\documentstyle[prl,aps,epsfig,preprint]{revtex}

\draft

\newlength{\textwidthm}
\begin{document}
\title{Dipole Blockade and Quantum Information 
Processing in Mesoscopic Atomic Ensembles}
\author{ M.~D.~Lukin$^1$, M.~Fleischhauer$^{1,2}$, R.Cote$^3$}
\address{$^{1}$ITAMP, Harvard-Smithsonian Center for Astrophysics,
                Cambridge, MA~~02138 }
\address{$^{2}$Fachbereich Physik, Universit\"at Kaiserslautern,
D-67663 Kaiserslautern, Germany}
\address{$^{3}$Physics Department, University of Connecticut, 
Storrs, Connecticut 06269}
\author{L.~M.~Duan, D.~Jaksch, J.~I.~Cirac and P.~Zoller}
 \address{Institut f\"ur Theoretische Physik, Universit\"at Innsbruck,
 A--6020 Innsbruck, Austria}

\date{\today}
\maketitle
\begin{abstract}
We describe a technique for manipulating  quantum 
information stored in collective states of mesoscopic ensembles. 
Quantum processing is accomplished by optical 
excitation into states with strong dipole-dipole interactions. 
The resulting ``dipole blockade'' can be used to inhibit 
transitions into all but singly excited collective states. 
This can be employed for a controlled generation of 
collective atomic spin states as well 
as non-classical photonic states and for 
scalable quantum logic gates. An 
example involving a cold Rydberg gas is analyzed.

\end{abstract}

\pacs{PACS numbers 03.67.-a, 42.50.-p, 73.23.-b}





Recent advances in quantum information science have opened a 
door for a number of fascinating potential applications ranging from 
the factorization of large numbers and secure communication to 
spectroscopic techniques with enhanced sensitivity. 
But the practical implementation of quantum processing protocols
such as quantum computation 
requires coherent manipulation of a large 
number of coupled quantum 
systems which is an extremely difficult task \cite{rev}.
Challenges ranging from  a long-time storage 
of quantum information to scalable quantum logic gates are by now well known. 
It is generally believed that precise manipulation of {\it microscopic} 
quantum objects is essential to implement quantum protocols. For example 
in most of the potentially viable candidates for quantum computers
an exceptional degree of control over sub-micron systems is 
essential for performing single-bit operations 
and the two-bit coupling is accomplished by interactions
between nearest neighbors \cite{atoms}. 
Related techniques are also being explored 
that involve photons to connect qubits \cite{pel}, and to construct 
potentially scalable quantum networks \cite{cirac}. However, 
since the single-atom absorption cross-section is very 
small, reliable coupling to light requires high-finesse micro-cavities 
\cite{qed}.

In the present Letter we describe a technique for 
the coherent manipulation of quantum information stored in collective 
excitations of mesoscopic many-atom ensembles. 
This is accomplished by optically 
exciting the ensemble into  states with a strong atom-atom 
interaction. Specifically, we consider the case involving dipole-dipole
interactions in an ensemble of cold atoms excited into Rydberg states.
Under certain conditions the level shifts associated with these interactions
can be used to block the transitions into states with more than a single
excitation.  The resulting ``dipole blockade'' phenomenon closely resembles 
similar mesoscopic effects in nanoscale solid-state devices \cite{block}. 
In the present
context it can take place in an ensemble with a size that can exceed 
many optical wavelengths. Combined with the
exceptional degree of control that is typical for quantum optical 
systems and long coherence times, this allows to considerably 
alleviate many stringent requirements for the experimental 
implementation of various quantum processing protocols. In particular, we show
that this technique can be used to i) generate superpositions
of collective spin states (or Dicke states) in an ensemble; 
ii) coherently convert these
states into  corresponding states of photon wavepackets of prescribed 
direction, duration 
and pulse-shapes and vice versa using the 
collectively enhanced coupling to light
\cite{memory1}; 
and iii) perform quantum gate operations
between distant qubits.  Corresponding applications including (i) sub-shot noise
spectroscopy and atom interferometry \cite{win}, (ii) secure 
cryptography protocols \cite{BB84}, 
and (iii) scalable quantum logic
devices can be foreseen.  In general, no  strongly coupling micro-cavities and 
no single particle control is required to implement computation and 
communication protocols.  
We further anticipate that the approach can be applied to a variety of 
interacting many-body systems ranging from 
trapped ions to specifically designed semiconductor structures.

The basic element of the present scheme is an ensemble
of $N$ identical multi-state atoms (Fig.1) contained in a volume $V$.
Using well-developed techniques all atoms  can be trapped and 
prepared in a specific sub-level ($|g_i\rangle$,  $i=1...N$) 
of the ground state manifold. 
Relevant states of each atom 
include: a pair of meta-stable sub-levels of the ground state 
manifold $|q_{i}\rangle$ that are used for long-time storage of qubits 
(storage states) and  long-lived Rydberg states 
$|r_{i} \rangle,  |p'_i\rangle, |p''_i\rangle$. Additional Rydberg
sublevels as well as lower electronic excited 
states can be used for specific applications. 
We assume modest atomic densities, such that
interactions between atoms can safely be neglected whenever
they are in the sub-levels of the ground state. This also implies 
long coherence lifetimes -- up to few seconds -- 
associated  with the storage sub-levels.
However when  excited into the Rydberg states the atoms interact 
strongly with each other due to the presence of resonant
long-wavelength dipole-dipole interactions.

The manipulation of atoms is done by light
fields of different frequencies and polarizations, which illuminate 
the entire ensemble and excite all atoms with equal 
probability. Throughout the paper we will assume that hyperfine sublevels
of the ground state and Rydberg states are excited in a  Doppler free 
configuration, which  can be done using two-photon processes.
The Hamiltonian describing these
interactions can be written in terms of collective operators  
$\hat\Sigma_{\mu\nu} = \sum_i |\mu_i\rangle\langle 
\nu_i| /\sqrt{N}$, where $|\mu_i\rangle, |\nu_i\rangle$ denote single-atom
eigenstates. As a consequence only
{\it symmetric collective states} are involved in the process. 
The relevant ones are here the ground state
$ |{\bf g}\rangle \equiv
|{\bf g}^N\rangle = |g_1\rangle ... |g_N\rangle$, the $n$-times
excited storage states $|{\bf q}^n\rangle \equiv 
|{\bf g}^{N-n},{\bf q}^n\rangle = \sqrt{(N-n)! N^n/(N! n!)}\, 
(\hat\Sigma_{gq}^{+})^n |{\bf g}\rangle$, and the 
Rydberg states $|{\bf r}^n\rangle \equiv 
|{\bf g}^{N-n},{\bf r}^n\rangle= \sqrt{(N-n)! N^n/(N! n!)}\, 
(\hat\Sigma_{gr}^+)^n |{\bf g}\rangle $.
For a sufficiently small number $n \ll N$, these
are bosonic excitations, 
since $[\hat\Sigma_{\mu\nu},\hat\Sigma_{\mu\nu}^+] \approx 1$.

We now discuss the interaction between collective atomic excitations.
We assume, for a moment, that the atoms are ``frozen'' in space.  
To be specific we here consider excitation 
``hopping'' via resonant dipole-dipole interactions  
between Rydberg 
atoms (F{\"o}ster process \cite{fost}) as illustrated in 
Fig.~1b \cite{coldrydb}. 
The energy separation between the optically excited Rydberg state 
$|r_i\rangle$ and the pair of the sub-levels of different parity 
$|p'_i\rangle, |p''_i\rangle$ is adjusted (using e.g. 
electric fields) such that $E_r - E_{p'} = 
E_{p''}- E_r$. In this case, any pair of atoms excited in the 
$|r_i\rangle|r_j\rangle$ 
states would undergo a hopping transition into the states 
$|p'_i\rangle |p''_j\rangle$ and $|p''_i\rangle |p'_j\rangle$
etc., 
resulting in a splitting of the excited Rydberg components. 
For a system with three effective Rydberg states (Fig.1b) the relevant 
process is  described by the Hamiltonian:

\begin{eqnarray}
{\hat V}_d = \hbar \sum_{i>j} \kappa_{ij}  
 |r_i\rangle |r_j\rangle \left( \langle p'_i| \langle p''_j| 
+ \langle p'_j| \langle p''_i| \right) + {\rm h.c.},
\label{ham1}
\end{eqnarray}
where 
$\hbar \kappa_{ij} \sim \wp_{rp'} \wp_{rp''}/r_{ij}^3$, 
$\wp_{kl}$ being the dipole 
matrix elements for the corresponding transitions and $r_{ij}$ is the 
distance between the two atoms \cite{sym}. 
In general this interaction does not affect the 
singly-excited collective states (e.g. ${\hat V}_d |{\bf r}\rangle = 0$) but
leads to a splitting 
of the levels when two or more atoms are excited. 
In particular, the collective eigenstate with atoms $i$ and $j$ being excited
and all other atoms in the ground or storage states
\begin{eqnarray}
|{\bf\pm_{ij}^2}\rangle &=& {1 \over \sqrt{2}} \Bigl[ |g_1 ... r_i ... r_j ...
g_N \rangle \\
&\pm& 
\bigl(| g_1 ... p'_i ... p''_j ... g_N\rangle 
+ | g_1 ... p'_j ...  p''_i ... g_N\rangle\bigr)/\sqrt{2} \Bigr] \nonumber  
\end{eqnarray}
are split by $\hbar \kappa_{ij}$. It follows that for an ensemble 
contained in a finite
volume $V$ the manifold of doubly excited states has 
a minimum splitting
of order $\bar\kappa= \wp_{rp'}\wp_{rp''}/(\hbar V)$ (Fig.2). 
Thus if $\bar\kappa$ is much larger than the
linewidth ${\bf \gamma}_r$ of the Rydberg state, 
resonant excitations from the singly to the doubly-excited
state are strongly suppressed. This is the essence of the dipole blockade.

Turning to the interaction of such many-atom systems with light 
we first discuss 
how to create single collective qubits. 
When the splittings of the doubly-excited
Rydberg states are sufficiently large,
a weak light field tuned to the single-atom resonance frequency can excite 
the transition
between the ground state and the first collective state $|{\bf r}\rangle$. 
But due to the splitting of the states $|{\bf \pm_{ij}}^2\rangle$ 
successive transitions into these and higher states are strongly inhibited. 
Hence if the atomic system is initially 
in its ground state $|{\bf g} \rangle$ the evolution is given by  
the effective two-level dynamics 
\begin{equation}
\left[\matrix{ |{\bf g}(t)\rangle \cr |{\bf r}(t)\rangle }\right]
=\left[\matrix{\cos\theta(t) & - i \sin\theta(t) \cr i 
\sin\theta(t) & \cos\theta(t)}
\right] \left[\matrix{ |{\bf g}(0)\rangle \cr |{\bf r}(0)\rangle}\right],
\label{colrabi}
\end{equation}
where $\theta(t) = \sqrt{N} \int_0^t \Omega(\tau)/2 d\tau$. 
The ensemble 
displays Rabi-oscillations only between ground and singly 
excited Rydberg state with a collective Rabi rate $\sqrt{N} \Omega$. 
There are obviously several conditions that restrict the  validity of 
Eq.(\ref{colrabi}). 
First of all, the Hamiltonian
evolution takes place for times $t$ much shorter than 
the dephasing time ${\bf \gamma_r}^{-1}$. 
Secondly, there is a finite  probability to populate the doubly excited
states $|{\bf \pm_{ij}},2\rangle$ that scales as $\sim \bar
\kappa^{-2}$. 
For example, at a time $t=T$ where a single collective excitation is generated 
(i.e. where $\theta(T) = \pi$) the probabilities of errors due to 
populating the doubly excited states and dephasing 
scale, respectively,  as 
\begin{equation}
p_{\rm doub} \sim  { 1\over N^2} \sum_{i.j} 
{1 \over \kappa_{ij}^2T^2} \approx { 1 \over 4\pi (\bar \kappa T)^{2}},
\quad p_{\rm deph} 
\sim {\bf \gamma_r} T \label{err}.
\end{equation}
Hence, the system can be driven into superpositions of 
collective states  $\alpha_0 |{\bf g} \rangle + \alpha_1 
|{\bf r}\rangle$.  The single-quantum 
excitation  
can now be stimulated into a storage 
sub-level (e.g. $|{\bf q}\rangle$) by a $\pi$-pulse ($\int\Omega_q(\tau)  
d \tau = \pi$) and we can associate a qubit with the 
state $ \alpha_0|{\bf g} \rangle+\alpha_1|{\bf q}\rangle$.  
As indicated in Fig.~2, this procedure can 
be generalized to the generation of higher order collective states
or their superpositions using a sequence of
properly timed pulses. For example, the sequence of two pulses $\sqrt{N-n} 
\int\Omega(\tau) d\tau = \pi$ and $\sqrt{n+1} 
\int\Omega_q(\tau) d\tau = \pi$ results in $|{\bf q}^n\rangle
\rightarrow |{\bf q}^{n+1}\rangle$. 
Note that under the conditions of dipole 
blockade the transitions 
into collective Rydberg states other than $|{\bf r},{\bf q}^m\rangle$
are always inhibited. 
The scaling of the error probability
in each step is given by Eq.(\ref{err}).
To prove that a 
synthesis of arbitrary superpositions $|{\bf \Psi}_n\rangle = 
\sum_{m=0}^n\alpha_m |{\bf q}^m\rangle$ is possible we note that 
the inverse procedure, i.e. 
the evolution from a known state $|{\bf \Psi}_n\rangle$ into the ground state 
$|{\bf g}\rangle$ can easily be constructed. 
In particular, $2n$ pulses of proper 
duration can be used to sequentially empty the states $|{\bf q}^n\rangle,
|{\bf r};{\bf q}^{n-1}\rangle, |{\bf q}^{n-1}\rangle $ etc, and the system
would evolve into its ground state. Since this evolution is unitary
a reverse sequence results in 
$|{\bf g} \rangle \rightarrow |{\bf \Psi}_n\rangle$.

Before proceeding we note several important properties of 
collective spin states generated by means of the dipole blockade.
First of all, these states
are precisely of the from required to maximize 
sensitivity of atomic clocks \cite{win} beyond the quantum limit. Secondly, 
by using standard techniques of atom optics it is possible to couple
the atoms in a specific internal state out of the trap. When these techniques 
are applied to the atoms in state $|q\rangle$, the
states with a desired number of atoms can be coupled out. 
In addition to a number of important potential applications for entanglement
engineering in atom interferometers, such an ``atom gun''  can make 
fascinating experiments involving spatially delocalized 
Schr{\"o}dinger cat states possible. Finally,   
the spin states  $|{\bf q}^n\rangle$ exhibit a collectively 
enhanced coupling to light, when excited in a proper two-photon configuration.
Consequently, the state $|{\bf \Psi}_n\rangle$ can be transfered from the spin
degrees of freedom to the optical field \cite{memory1}. If the optical 
thickness of the cloud at the resonance wavelength is sufficiently high, this
allows to generate wavepackets
on demand with desired properties such as quantum states, pulse shapes 
and duration without the use of high-Q cavities as necessary in  
single-atom cavity-QED \cite{photon-gun}. 
The quantum state of these wavepackets can be transferred to 
distant ensembles \cite{memory1} thereby making quantum networking 
possible.

The dipole blockade mechanism also facilitates quantum logic operations.
For instance, in  
proof-of-principle experiments qubits stored in few-$\mu m$ spaced 
atomic clouds can be entangled if the transitions in one ensemble 
are inhibited when a collective Rydberg state is excited in a second 
ensemble. In essence, this scheme utilizes the same principle as that of 
the Ref.\cite{singlerydb} with the single-particle excitations 
replaced by collective
qubits stored e.g. in $ \bigl\{|{\bf g} \rangle, |{\bf q}\rangle \bigr\}$.  
In what follows we discuss another approach that allows for such
operations between distant ensembles, which is instrumental for
scalability. 
In this approach the qubit states of two 
different ensembles $ \bigl\{|{\bf g} \rangle_1,|{\bf q}\rangle_1\bigr\}$
and $ \bigl\{|{\bf g} \rangle_2,|{\bf q}\rangle_2\bigr\}$
are first transferred to two hyperfine sublevels 
of  one  ensemble (as shown in Fig.3) via mapping to light fields:  
$|{\bf q}
\rangle_1 \rightarrow |{\bf q}_{+}
\rangle, |{\bf q}
\rangle_2 \rightarrow |{\bf q}_{-}
\rangle$. 
A universal logic gate is then performed in three steps. 
First, all atoms in the storage states $|q_{-}\rangle$ are 
excited into a processing Rydberg state $|r_{-}\rangle$
by a $\pi$-pulse,  $\int_0^T  
{\Omega_{-}}(\tau)d\tau = \pi$. Second, the atoms in the storage states 
$|q_{+}\rangle$ are 
excited into a processing state $|r_{+i}\rangle$ and back  
by a $2\pi$-pulse,  $\int_0^T  
{\Omega_{+}}(\tau)d\tau = 2 \pi$. This results in a phase shift of $\pi$
for the collective state $|{\bf q_+}\rangle$. 
Note, however, that transitions into the doubly excited states
$|g_1 ... r_{+i} ... r_{-j} ... g_N\rangle_1$ are inhibited due to 
the dipole blockade. 
As a result, the atoms in the 
collective state $|{\bf r_-,q_+}\rangle$ do not acquire any phase rotation. 
In a final step, the atoms in state $|r_{-}\rangle$ are stimulated back 
into the
storage sublevels resulting in a total phase shift of $\pi$ for the  
states
$|{\bf q_\pm}\rangle$ and $|{\bf q_+,q_-}\rangle$. Hence, a conditional
phase shift
\begin{eqnarray}
|{\bf g}\rangle \rightarrow |{\bf g}\rangle, \;
|{\bf q_{\pm}}\rangle \rightarrow - |{\bf q_{\pm}}\rangle, \;
|{\bf q}_{+},{\bf q}_{-}\rangle 
\rightarrow - |{\bf q}_{+},{\bf q}_{-}\rangle 
\label{phase}
\end{eqnarray} 
is obtained. The gate is completed by 
sending the qubits back 
to the original registers. The leading 
sources of errors are again due to the  finite probability of populating 
the doubly excited states $p_{doub}$, and due to dephasing of Rydberg 
transitions $p_{deph}$, given by Eq.(\ref{err}).

We next discuss a number of important practical 
issues associated with the present technique.
First of all the decoherence of collective  
excitations is of particular importance here: the qubit states
such as $|{\bf q_\pm}\rangle, |{\bf r}\rangle$ are, 
in fact,  entangled  superpositions 
of $N$ single-particle states and these might be expected to be extremely 
fragile. This is not the case, since the symmetric entangled states 
involved are known to be very robust against e.g. particle loss 
\cite{cirac2000}. As a result, 
collective dephasing rates 
are equal to the corresponding single-particle rates,
if the dephasing is dominated by single-particle rather than collective 
processes. The last assumption is well justified as long as 
the average inter-atomic distance 
is below the reduced optical wavelength $\lambda/(2\pi)$: 
$N/V \le (2\pi/\lambda)^3$ \cite{note2}. Errors due to atom-number 
fluctuations ($\Delta N$) are negligible, provided that 
$\Delta N \ll N$. Finally, because the states that 
are involved in strong atomic interactions are, in the ideal limit,
never populated, the present approach avoids mechanical interaction between 
atoms and leaves the qubit degrees of freedom decoupled from 
atomic motion \cite{singlerydb}.  

Present day magnetic or optical traps allow one to contain about 
$N \sim 10^4$  cold alkali atoms (Li, Cs or Rb) within $V < ( 10 \mu 
{\rm m})^3$. For the atoms excited to Rydberg 
states ($n\sim 50$), typical coherence 
lifetimes limited by radiative relaxation and 
black-body radiation  correspond to 
${\bf \gamma_r} < 10$kHz, whereas dipole-dipole interactions 
correspond  to  values of ${\bar\kappa}$ in the range of 10-100 MHz. This 
implies that for excitation on a fast time scale
such as  $T < 100$ ns,
the probability of loss ${p}_{doub},p_{loss} < 1 \%$. 
Such  ensembles are opaque near the resonant optical transitions of the 
$D_{1,2}$ absorption lines, thereby making efficient coupling to light 
possible.  In a potentially scalable approach 
one could employ 
arrays of such traps created on nanofabricated 
surfaces \cite{traps}, in which case the traps could be also
coupled using low-Q cavities and low-loss fiber waveguides.  
Finally, since the 
atomic motion
is never coupled to the qubits, the temperature is only limited by the 
requirement that the atomic distribution should not change significantly 
on the time  scale of the gate operation $T$
(``frozen'' gas approximation). This is the case even for
temperatures as high as a few $mK$.

In summary, we have shown that the dipole blockade can be used 
for coherent manipulation and entanglement of collective excitations in
mesoscopic ensembles of cold atoms. We anticipate a number of 
potential applications ranging from atomic clocks and secure quantum 
communication to quantum computation with mesoscopic atomic samples.

We thank E.~Eyler, P.~Gould, T.~Gallagher and V.~Kharchenko for many 
stimulating discussions. 
This work was supported by the Austrian Science Foundation, the Europe Union
project EQUIP, the ESF, the European TMR network Quantum Information,  
the NSF through the grant to the ITAMP and the NSF ITR program. M.F. 
thanks the Deutsche Forschungsgemeinschaft for
financial support and ITAMP for the hospitality during his stay.

\def\etal{\textit{et al.}}


\begin{figure}[ht]
\centerline{\epsfig{file=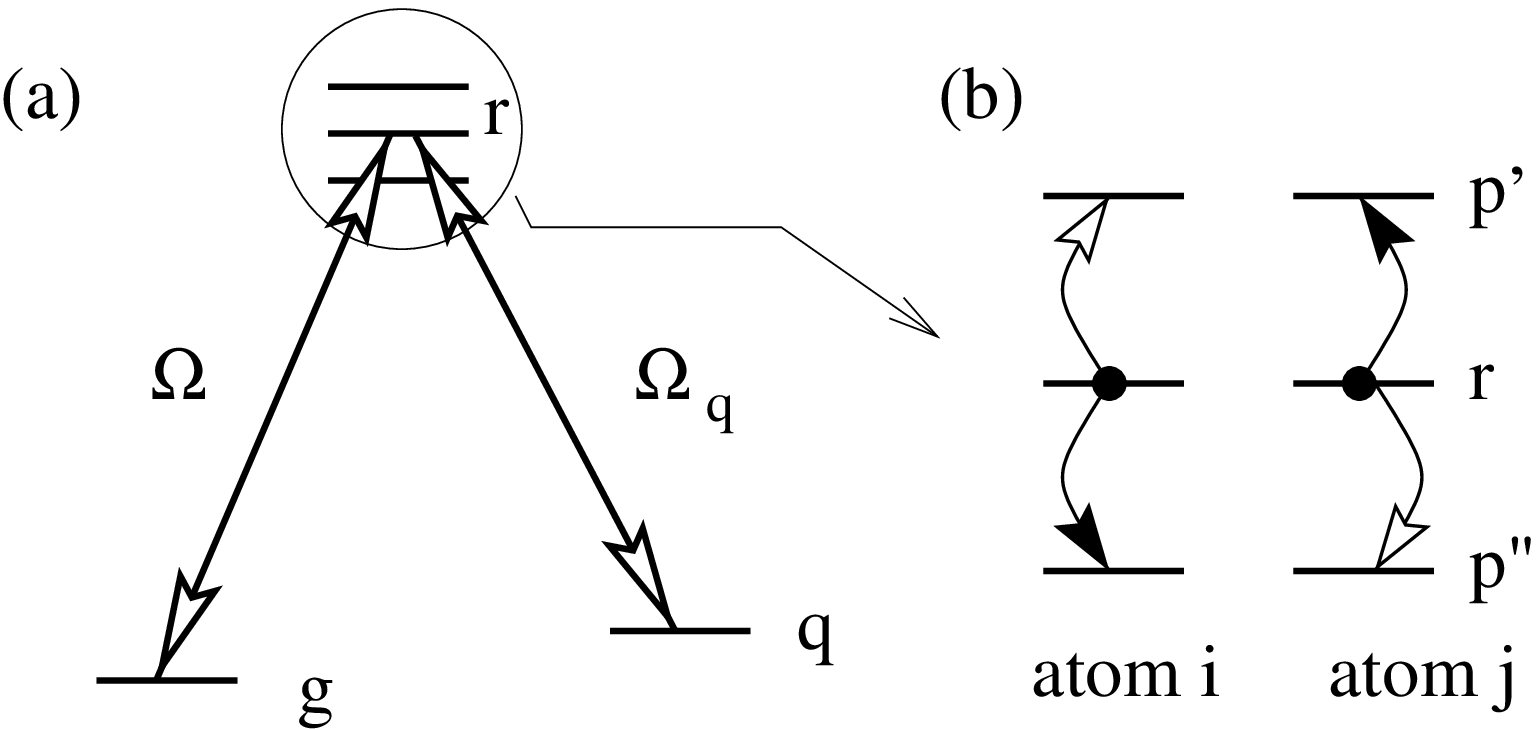,width=8 cm}}
 \vspace*{2ex}
 \caption{ (a) Relevant single-atom levels for quantum information 
storage ($q $) and processing ($r,p',p''$) can be accessed by 
optical fields. $\Omega,\Omega_q$ are Rabi-frequencies. (b) 
Resonant atom-atom interaction causes 
excitation hopping between 
atom pairs and results in collective level splitting. Arrows indicate 
atomic transitions in the hopping process.
}  
\end{figure}



\begin{figure}[ht]
\centerline{\epsfig{file=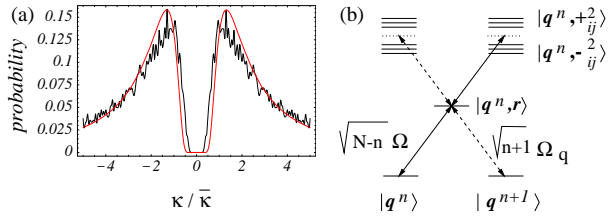,width=8 cm}}
 \vspace*{2ex}
\caption{ 
(a) Probability distribution for frequency splittings of manifold
of doubly-excited Rydberg states. Shown is a
Monte-Carlo simulation for $3\times 10^4$ positions
in a rectangular box and analytical  approximation for a random gas
$p(x)=\sqrt{2}\pi \exp[-\pi^3/18 x^2]/ (6 x^2)$, $x = \kappa/{\bar \kappa}$.
(b) Successive generation of Fock states $|{\bf q}^n\rangle$
using dipole-blockade.
}
\label{splitting}
\end{figure}



\begin{figure}[ht]
\centerline{\epsfig{file=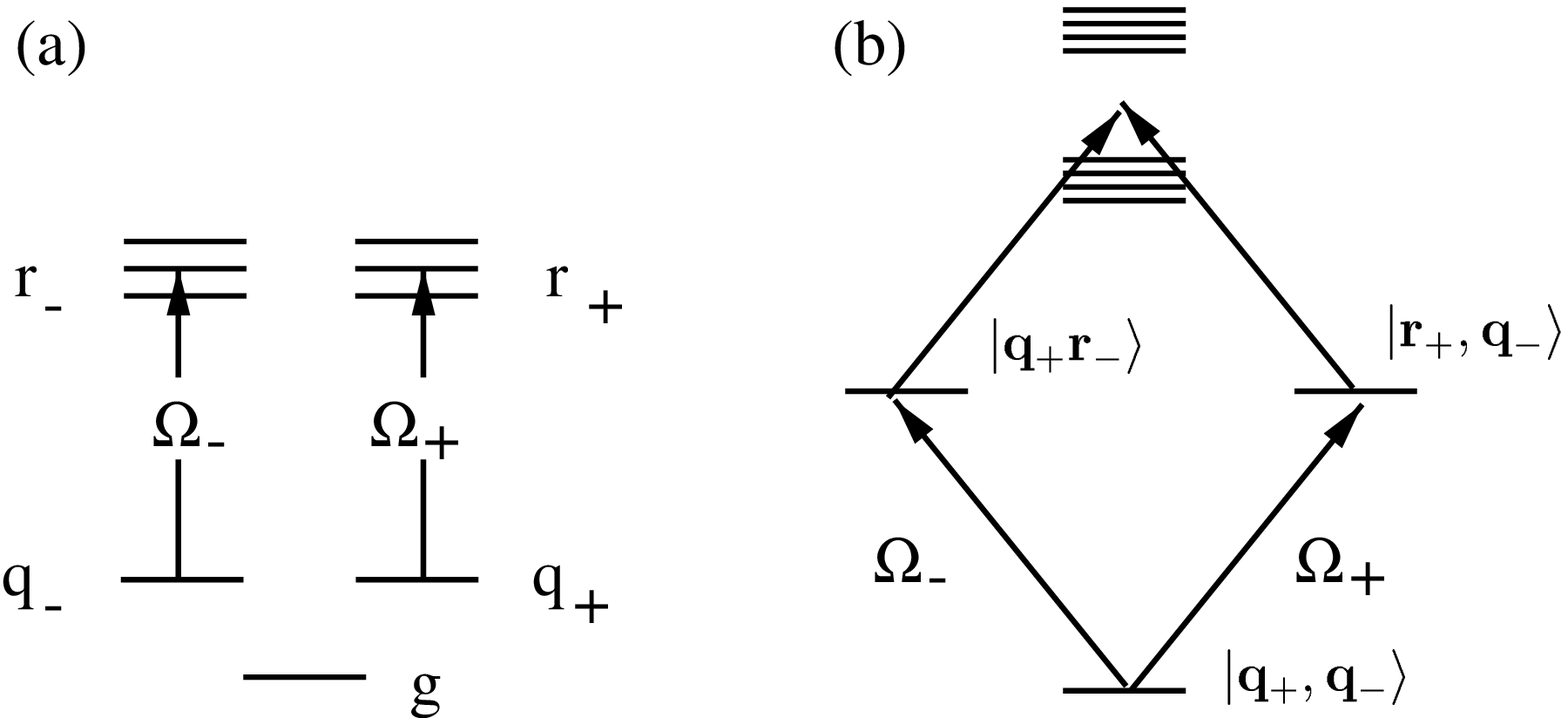,width=8cm}}
 \vspace*{2ex}
 \caption{(a) Level scheme for generation of conditional phase shifts. 
(b) Coupling scheme within the manifold of doubly excited collective  states. }
\label{gate}
\end{figure}



\begin{thebibliography}{99}


\bibitem{rev} 
See e.g.:
D. P. DiVincenzo, Science {\bf 270}, 255 (1995);
C. H. Bennett, Phys. Today {\bf 48}, No. 10, 24 (1995);
A. Ekert and R. Josza, Rev. Mod. Phys. {\bf 68}, 733 (1996). 


\bibitem{atoms} See
J. Gruska, Quantum Computing (McGraw-Hill, London, 1999);
A. Steane, Rep. Prog. Phys. {\bf 61}, 117 (1998).

\bibitem{pel} 
T.~Pelizzari, S.~A.~Gardiner, J.~I.~Cirac, and P.~Zoller,
Phys.Rev.Lett. {\bf 75}, 3788 (1995).

\bibitem{cirac} J.~I.~Cirac, P.~Zoller, H.~Mabuchi and H.~J.~Kimble, 
Phys.Rev.Lett. {\bf 78}, 3221 (1997). 


\bibitem{qed} Q.A.Turchette {\it et al.}, Phys.Rev.Lett. {\bf 75}, 4710 
(1995). 

\bibitem{block} See e.g. {\it Mesoscopic phenomena in Solids}, B.L.Altshuler, 
P.A.Lee, and R.Webb, Eds., (Elsevier,Amsterdam, 1991). 


\bibitem{win}  D.J. Wineland, {\it et al.}, Phys.Rev.A {\bf 46}, R6797 (1992); 
J. Bollinger {\it et al}., 
{\it ibid} {\bf 54}, 4649 (1996). S. F. Huelga, {\it et al.} Phys. Rev. Lett.
{\bf 79}, 3865 (1997);
P.~Boyer and M.~A.~Kasevich, Phys.~Rev.~A {\bf 56}, R1083 (1997).

\bibitem{BB84} Deterministic single-photon sources are crucial for physical
implementation of a secure BB84 quantum cryptography protocol; see. e.g. 
N. Luetkenhaus, Phys.~Rev.~ A, {\bf 61}, 052304 (2000).


\bibitem{memory1}   
M.~D.~Lukin, S.~F.~Yelin, and M.~Fleischhauer, Phys.~Rev.~Lett., 
{\bf 84}, 4232 (2000);
M.~Fleischhauer and M.~D.~Lukin, Phys.~Rev.~Lett.  {\bf 84}, 5094 (2000);
L.M. Duan, J.I. Cirac, and P. Zoller, unpublished.

\bibitem{galla} 
T.F.~Gallagher, {\em Rydberg atoms}, (Cambridge University Press New York 
1994).

\bibitem{fost} 
Th. F{\"o}ster, in {\it Modern Quantum Chemistry},
O.~Sinanoglu Ed., (Academic Press, New-York, 1996).

\bibitem{coldrydb} Such processes have been observed
in a cold Rydberg matter, see e.g. 
I.~Mourachko, et al., Phys.~Rev.~Lett.~{\bf 80} 253 (1998); W.R.~Anderson 
et al., {\it ibid.}, 249.

\bibitem{sym} E.g. in a case when the Rydberg $S$-state  is excited by light 
the pair-coupling strength $\kappa_{ij}$ depends only on the separation of the
atoms for symmetry reasons.

\bibitem{photon-gun} F.~De Martini {\it et al.}, Phys. Rev. Lett.
{\bf 76}, 900 (1996); C.~K.~Law and H.~J.~Kimble, J. Mod. Opt.{\bf 44}, 
2067 (1997); M. Hennrich {\it et al.}, Phys.~Rev.~Lett, in press.

\bibitem{singlerydb} These issues have been discussed in detail for 
quantum gates based on individual Rydberg atoms, 
D. Jaksch {\it et al}, Phys.~Rev.~Lett.~{\bf 85} 2208 (2000). 

\bibitem{cirac2000} W. D\"ur, G. Vidal and J. I. Cirac, quant-ph/0005115.

\bibitem{note2} Levels can also be shifted due to interaction of the 
excited Rydberg atoms with ground state atoms. The magnitude of these shifts 
is at most few hundred KHz at densities $N/V \sim 10^{13}$ cm$^{-3}$. These 
shifts can be compensated by a small detuning since  
the broadening is expected to be much smaller than the shift 
at low temperatures.   

\bibitem{traps} J.~Reichel, W.~H{\"a}nsel, and T.~W.~H{\"a}nsch , 
Phys.~Rev.~Lett.~{\bf 83} 3398 (1999);
R.Folman {\it et al.}, Phys.~Rev.~Lett.~{\bf 84} 4749 (2000).

\end{thebibliography}
\end{document}